\documentclass{aa}

\usepackage{graphics}
\begin{document}

   \title{Kinematics and dynamics of the ``superthin'' edge-on disk galaxy 
          IC 5249}

   \author{P.C. van der Kruit 
   \inst{1}
   \and 
   J. Jim\'enez-Vicente 
   \inst{1}
   \and 
   M. Kregel 
   \inst{1}
   \and 
   K.C. Freeman 
   \inst{2}
   }

   \offprints{P.C. van der Kruit}

   \institute{Kapteyn Astronomical Institute, University of Groningen,
   P.O. Box 800, 9700 AV Groningen, the Netherlands\\
   \email{vdkruit@astro.rug.nl, jjimenez@astro.rug.nl, kregel@astro.rug.nl}
   \and
   Research School of Astronomy and Astrophysics,
   Mount Stromlo and Siding Spring Observatories, Australian National 
   University, Private Bag, Weston Creek, 2611 Canberra, Australia\\
   \email{kcf@mso.anu.edu.au}
   }

   \date{Received ; accepted }

   \maketitle

   \abstract{
We present spectroscopic observations of the stellar motions in the disk
of the superthin edge-on spiral galaxy IC 5249 and re-analyse 
synthesis observations of the HI.\\ 
We find that the HI rotation curve
rises initially to about 90-100 km/s, but contrary to the conclusion of Abe
et al. (1999) flattens well before the edge of the optical disk. Over most
part of the optical disk we have been able to establish that the
(tangential) stellar velocity dispersion is 25-30 km/s. We argue that
the central light concentration in the disk is not a bulge in the
classical Population II sense, but most likely represents structure in
the disk component. From earlier surface photometry we adopt a value for
the radial scalelength of the disk of $7 \pm 1$ kpc, a vertical
scaleheight of $0.65 \pm 0.05$ kpc and a disk truncation radius of $17
\pm 1$ kpc. The HI disk has a measurable thickness but from our analysis
we conclude that this is due to a small inclination away from
perfectly edge-on.\\
The very thin appearance of IC 5249 on the sky is the result of a
combination of a low (face-on) surface brightness, a long scalelength
and a a sharp truncation at only about 2.5 scalelengths. In terms of the
ratio of the radial scalelength and the vertical scaleheight of the
disk, IC 5249 is not very flat; in fact it is slightly fatter than the
disk of our Galaxy. From various arguments we derive the stellar 
velocity dispersions at a position  one radial scalelength
out in the disk ($R \sim 7$ kpc) as respectively
$\sigma _{\rm R} \sim $ 35 km/s, $\sigma _{\theta} \sim $ 30 km/s and 
$\sigma _{\rm z} \sim $ 20 km/s. This is comparable to the values for
the disk of our Galaxy in the solar neighborhood. Near the edge of the
disk the ratio of radial to vertical velocity dispersion is probably higher.\\
Presumably the angular momentum distribution of the gas that formed the disk in
IC 5249 was such that, compared to the Galaxy, 
a much more extended distribution resulted in spite of
the lower overall rotation and mass. The low surface density that arose from 
that resulted in a thicker HI layer in which star formation proceeded at
a much slower rate, but disk heating proceeded at a similar pace. 
       }        
       
      
%

\section{Introduction}

The stellar disks of spiral galaxies have a thickness that is thought to
arise from the secular evolution of the random velocities of the stars.
In the course of time the vertical 
(and other) velocity dispersions increase (but level
off when the stars spend most of their time outside the gas-layer where much
of the vertical scattering is thought to occur) and consequently the
stellar disk thickens. When for the Galaxy we consider the solar neighborhood
we find that the $z$-velocity dispersion of the old disk stars is of order
20 km/s (e.g. Dehnen \&\ Binney, 1998),
 while the vertical distribution of the old disk stars can
be described (at least at larger distances $z$) by an exponential with
an $e$-folding (the scaleheight over which it drops by a factor $e$) 
of about 0.325 kpc (Gilmore \&\ Reid, 1983). 
On the other hand, the radial distribution of light in galaxy disks is 
also exponential but with a much larger e-folding (now called the 
scalelength). For the Galaxy the value of the disk scalelength is under 
dispute, but is in the range of 2.5 to 4.5 kpc (for a recent discussion 
see van der Kruit, 2000). For the majority of galaxies the ratio of the
two scale parameters is in the range 5 to 10 (van der Kruit \&\ Searle, 1981, 
1982; de Grijs, 1998, Kregel et al., 2001).

The random velocities of the stars in a disk are anisotropic; e.g. in 
the solar neighborhood the radial, tangential and vertical dispersions
are in the ratio 2.2 to 1.4 to 1.0 (Dehnen \&\ Binney, 1998). This ratio
contains information on the processes by which stars are being scattered
(Jenkins \& Binney, 1990; Jenkins, 1992). Furthermore, while the scaleheight
and scalelength of a disk are presumably determined independently (the 
scalelength from the angular momentum distribution in whatever the galaxy 
formed out of and the scaleheight as a result of secular evolution of stellar
motions) the radial-to-vertical axis ratio of the velocity ellipsoid is related
to the flattening of the stellar disks (van der Kruit \&\ de Grijs, 1999), 
because the radial velocity
dispersion can be related to the disk scalelength through Toomre's (1964)
criterion for local stability $Q$\lower.5ex\hbox{$\; \buildrel > \over 
\sim \;$}1 (for the definition of $Q$ see eq. 17 below). In fact,
if $Q \sim 2$ as is often found in numerical experiments, eq. 17 in van
der Kruit \&\ de Grijs (1999) gives a ratio of vertical to radial velocity
dispersion from 0.62 to 0.88 for the scalelength to scaleheight ratio in the 
range 10 to 5. 

\begin{figure}
{\resizebox{\hsize}{!}{\includegraphics{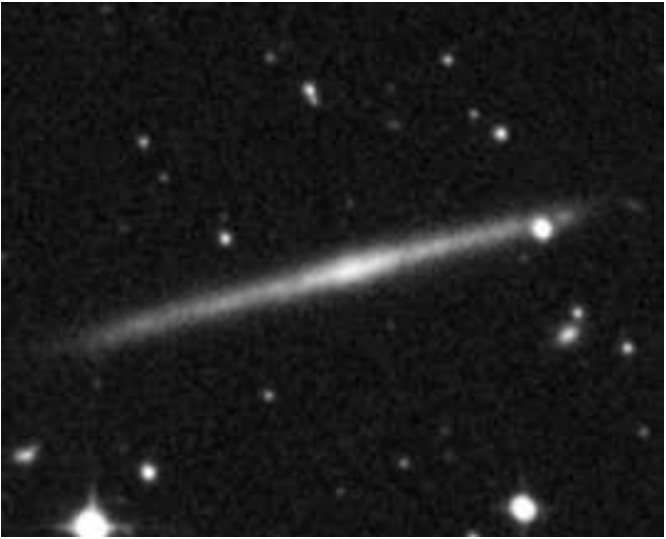}}}
\caption{Optical picture of IC 5249 taken from the second generation
red Digital Sky Survey. The frame measures 4.4 $\times $ 3.6 arcmin and
south is to the left.
}
\label{picture}
\end{figure}

Although the most common flattening of edge-on galaxies on the sky is
of order 5 to 10 in the ratio of the scale parameters and consequently also
in observed isophotes, there remains a class of so-called ``superthin''
galaxies, in which the axis ratio on the sky surveys appears much larger. 
A beautiful example of such a system is IC 5249, which has an axis ratio
of about 20 on the {\it Digital Sky Survey} (see Fig.~\ref{picture}). 
If indeed the ratio of the scalelength and the scaleheight would be that
large, the axis ratio of the stellar velocity ellipsoid can still be of
the order 0.45 according to the equation referred to above, but it would
seriously constrain the models for secular evolution that would have to be
much more efficient in the radial than in the vertical direction.

Photometry of IC 5249 has been performed by Carignan (1983) in the optical 
($B$-band) and by Wainscoat (1986) in the near-infrared ($H$, $J$ and $K$).
Carignan interpreted his photometry as that IC 5249 has two distinct
components, an inner one with a 18 kpc scalelength and an outer with 2.5
kpc, which suggest a long scalength disk with a shallow cut-off. Wainscoat's
photometry confirmed this and showed that the optical structure 
is not a result of
absorption by dust. He concluded that the flatness on the sky results
from the long apparent scalelength, since the vertical scaleheight was
not unusually small. 

More detailed photometry was published by 
Byun (1998) in $B$, $R$ and $I$ (and also
for ESO 404-G18). The disks turned out to have quite normal (even somewhat 
large) scaleheights, but exceptionally large scalelengths accompanied with
relatively sharp disk truncations. Also the photometry indicated that these are
examples of Low-Surface Brightness (LSB) galaxies seen edge-on. Further
photometry has been published by Abe et al. (1999) who drew attention
to a very sharp truncation of the stellar disk at less than two scalelengths.
Together with the apparent LSB-nature of the galaxy this gives rise to a
appearance on the sky of a very high axis ratio disk (see van der Kruit,
1999).

In the mean time more ``superthin'' galaxies have been studied, such as
UGC 7321 (Matthews et al., 1999; Matthews, 2000) and UGC 711 (Mendelowitz,
2000). Especially the thorough study of UGC 7321 by Matthews and co-workers
is interesting in that it gives values for the ratio of the scale parameters
from 14 in the inner regions to 10 further out. It also provides the smallest
measured value for a scaleheight (140-150 pc) in a stellar disk. 
UGC 7321 appears to be a superthin galaxy in the real sense. The radial 
scalelength is 2 kpc at the reddest wavelength measured ($H$) and the disk 
shows a truncation feature at about 7 kpc. We will return to the differences
and similarities between UGC 7321 and IC 5249 below.

In this paper we re-analyse the HI data used by Abe et al. (1999) in more
detail. We also present measurements of the stellar motions from a long
slit spectrum along the disk of the galaxy. In this paper we will 
use a distance of 36 Mpc, based on the radial velocity of about 2360 km/s.
At this distance 1 arcmin corresponds to 10.47 kpc. 
The appearance of IC 5249 and surface brightness contour maps (Byun, 1998)
show a central bright area. Byun interpreted this as a weak bulge. It
also appears to coincide with the rotation center of the galaxy. Unless
stated otherwise we have used this position as that of the center of
the galaxy. We will discuss the nature of this central area below.

\begin{figure}
{\resizebox{\hsize}{!}{\includegraphics{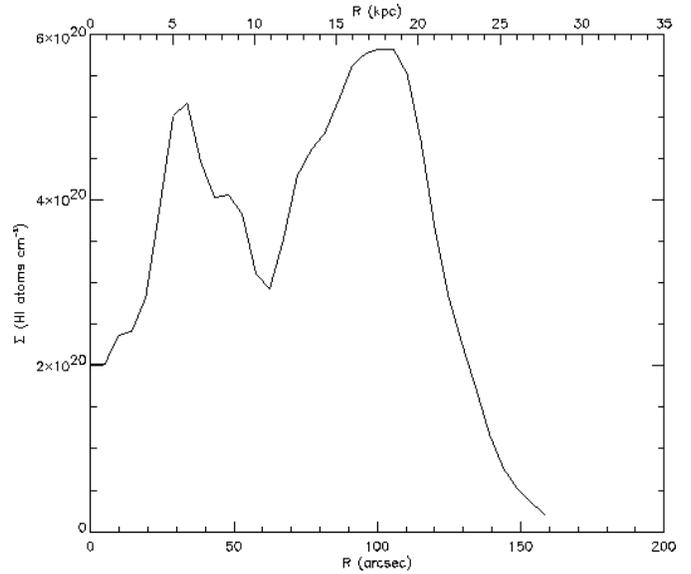}}}
\caption{Radial HI distribution from a decomposition of the observed
surface density projected onto the major axis. 
}
\label{radial HI}
\end{figure}

\section{The ATCA HI observations}

The observations used were the ones presented in Abe et al. (1999). They 
have been obtained on October 18, 1992 by P. Levasseur, C. Carignan and 
Y.-I. Byun with the Australia Telescope Compact Array (ATCA) 
during about 12 hours and were extracted for us from the ATCA
public archive and made available for
analysis. We were particularly interested in these observations for two
reasons. The first was the fact that we had some doubt about the analysis
in Abe et al. concerning the derived rotation curve (van der Kruit,
1999). Although realising the pitfalls (and describing these in their footnote
16) they derived a rotation curve from 
a first moment analysis in velocity at positions along the major axis. The
integrated profile (Mathewson et al., 1992) with a clear two-horned signature
and the published HI map by Abe et al. with fairly 
constant HI surface density on the sky
along the plane seemed inconsistent with their monotonically
rising rotation curve. We have therefore used the achival data to fully
analyse the position - velocity diagram. Our second interest was to see
whether the HI disk was resolved in the $z$-direction or whether useful limits
could be obtained from the data.

\begin{figure}
{\resizebox{\hsize}{!}{\includegraphics{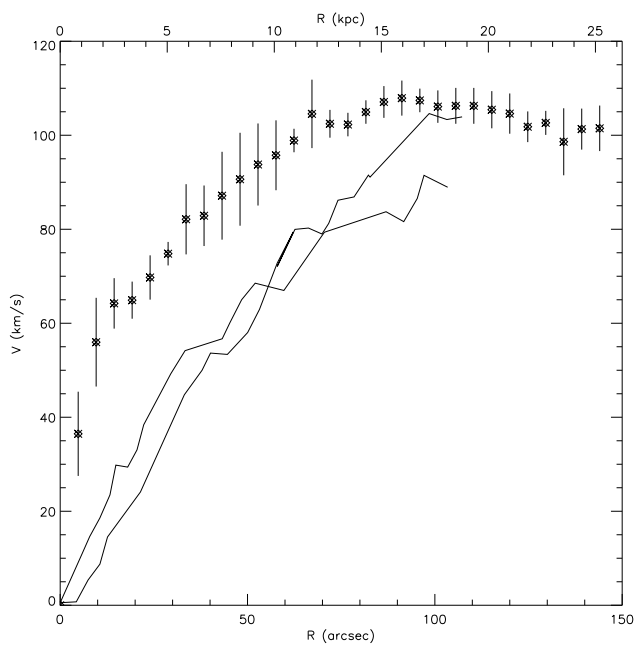}}}
\caption{The rotation curve of IC 5249 from a modeling of the distribution
of the HI in the position - velocity diagram. The rotation curve derived 
by Abe et al. (1999) from the same observations has been shown for 
comparison.  The beamsize along the major axis is 25\arcsec\ (FWHM).
}
\label{rotcurve}
\end{figure}

This analysis proceeded along the usual lines. First the data were transformed
into a single $(x,V)$-diagram ($x$ measured along the disk, $V$ the observed
radial velocity) by intergrating at each $x$ over all $z$. Then the observed
profile of HI surface density with distance from the center was unfolded 
into a radial HI surface density profile under the assumption of circular
symmetry. Before this both sides of the galaxy were averaged. The resulting
density profile is shown in Fig.~\ref{radial HI}. 
As noted by Abe et al., the ATCA
data add up to only $6 \times 10^{9}\ M_{\odot}$ (20 Jy km/s), while the
single dish measurements of Mathewson et al.(1992) imply $8.5 \times 10^{9}\ 
M_{\odot}$ (27.85 Jy km/s). It is possible that some emission is missing
as a result of the absense of short interferometer spacings. This then must be
emission over relatively large angular scale.

\begin{figure*}
{\resizebox{\hsize}{!}{\includegraphics{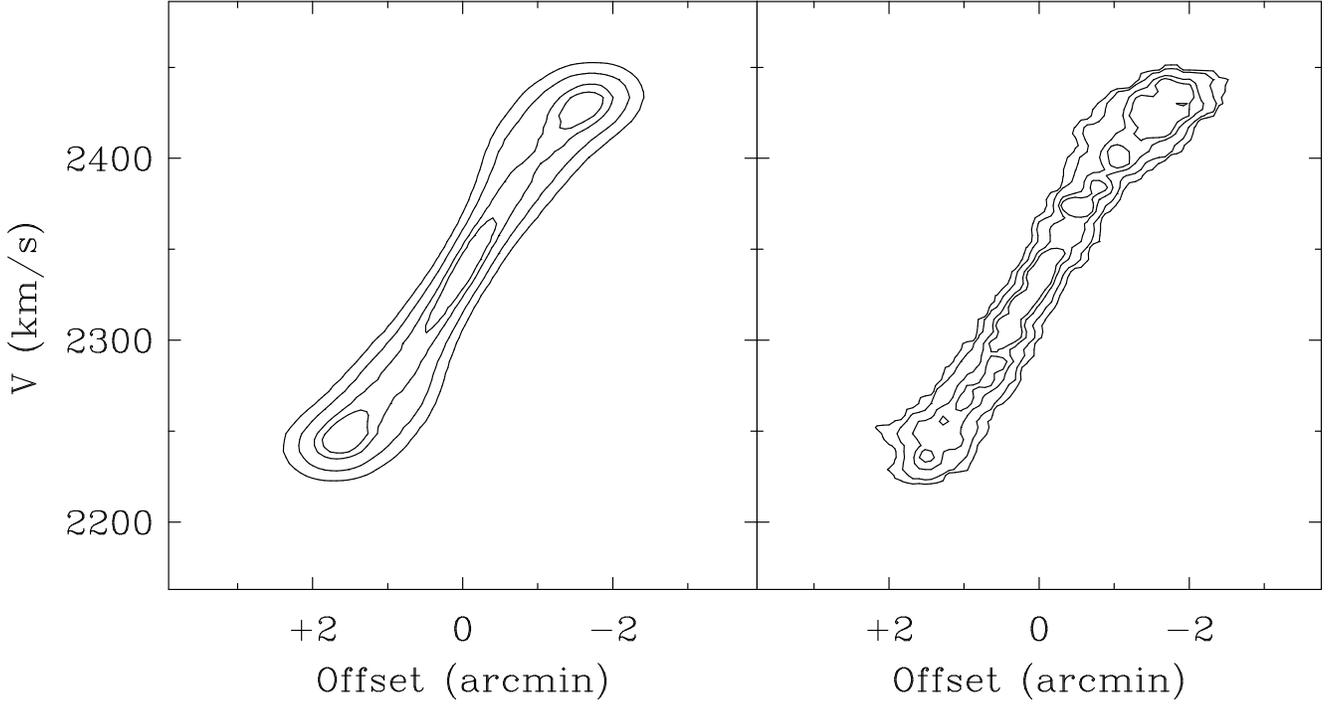}}}
\caption{Comparison of observed and calculated $(x,V)$-diagrams. The lefthand
panel shows the model and the righthand one the observations. The offset
$x$ is measured from the center. The contour levels correspond to 1.5, 3, 4.5
and 6 times the r.m.s. noise.
}
\label{xvdiagram}
\end{figure*}

\begin{figure}
{\resizebox{\hsize}{!}{\includegraphics{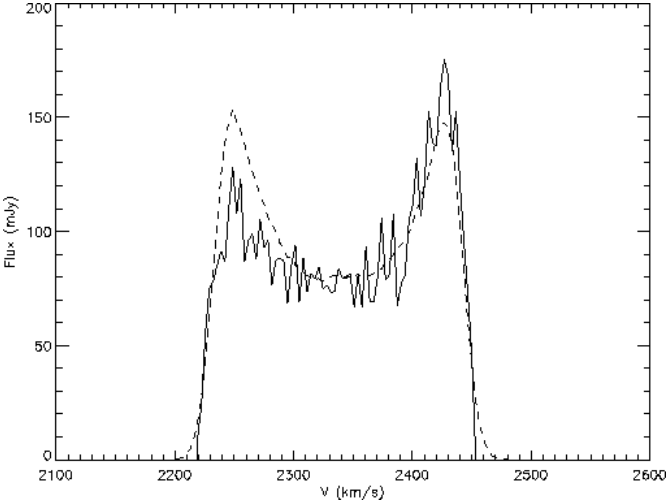}}}
\caption{Comparison of observed and calculated integrated HI profile.
}
\label{integralprofile}
\end{figure}

Next the $(x,V)$-diagram was used to make a first estimate for the rotation
curve and this was then used with the distribution in Fig.~\ref{radial HI} 
to derive
the corresponding distribution of HI emission in the $(x,V)$-diagram. 
In order to improve the sensitivity without degrading the quality of
the result, we have first smoothed the data
to a resolution (FWHM of the beam) of 20.20\arcsec$\times $26.18\arcsec. 
Also the
resulting integrated HI profile was generated and compared to the observed
one (in the synthesis observations). This was repeated with an improved 
estimate of the rotation curve until the best correspondence between model 
and observations was achieved. This resulted in the rotation curve
of Fig.~\ref{rotcurve}. 
The corresponding observed and calculated distribution
in the $(x,V)$-diagram are compared in Fig.~\ref{xvdiagram}, 
while Fig.~\ref{integralprofile} shows the 
observed and calculated integrated profiles. The procedure was repeated for 
a few values for the velocity dispersion of the HI. The best fit of the
data were obtained with an HI velocity dispersion of 7 km/s. 
A value of 8 km/s would also be consistent with the observations, but
10 km/s gave a significantly worse fit.

It can be seen that our rotation curve rises faster than the one by Abe et 
al., in broad terms reaching a flat part at some 105 km/s at about 60\arcsec\
($\sim $10 kpc) out to the edge of the observed HI at about 120\arcsec\
($\sim $21 kpc). The rotation curve derived 
by Abe et al. rises linearly from the center to about 100 km/s at 100\arcsec\ 
($\sim 17$ kpc). Along the major axis the beam (FWHM) is about 25\arcsec,
so the initial rise within 5 kpc may be affected by beam-smearing, since 
there the velocity gradient across the beam is appreciable.

\begin{figure}
{\resizebox{\hsize}{!}{\includegraphics{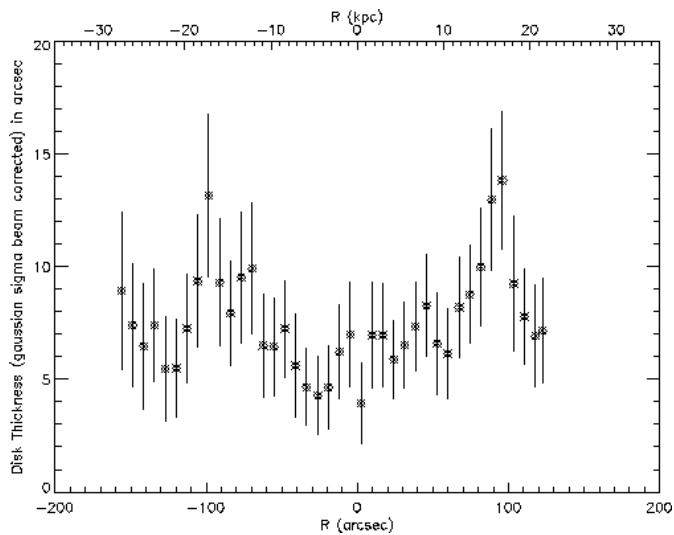}}}
\caption{The apparent thickness of the HI layer of IC 5249 with 
distance form the center. These measurements were obtained from the
total HI distribution as observed on the sky. The beamsize along the major
axis is 25\arcsec. 
}
\label{thickness}
\end{figure}

Our second concern was the thickness of the HI layer. In order to investigate
this we of course used high-resolution data, which means a beam of
10.9\arcsec$\times $12.5\arcsec\ (FWHM). 
In the direction perpendicular to the
disk the beam is then 12.4\arcsec. This is slightly larger then the full
resolution data would allow, but the latter were too noisy for a reliable
determination. The HI data were then fitted to Gaussians in the
$z$-direction and these were then corrected for the beam. The result of this is
in Fig.~\ref{thickness}.

The disk is clearly resolved in the $z$-direction. When fitted with
Gaussians we find that the
average Gaussian dispersion of the HI layer is $7.5 \pm 2.5$\arcsec\  
($1.3 \pm 0.4$ kpc). Over most part of the optical disk the HI layer has a 
reasonably constant thickness (again Gaussian dispersion) with a mean and 
r.m.s. scatter of $6.4 \pm 1.4$\arcsec\  
($1.1 \pm 0.3$ kpc). This is remarkably thick for a gas layer and 
comparable to the stellar disk! It is of course possible that
the HI consists of two components with different $z$-distributions (and
also velocity dispersions!), but we feel we would have seen the latter
in our analysis of the $(x,V)$-diagram since a more extended component
would (if substantially contributing to the observations) 
probably have had velocity dispersion larger than the canonical 7 km/s.

It should be noted, however, that the observed thickness may be the result of
a small inclination away from perfectly edge-on. We therefore also looked 
at the individual channel maps, but because of the lower
signal to noise we have not
been able to find any convincing indications for a measurable thickness
of the HI disk in those data. A further test would be to look at HI profiles
in the disk and just below/above this. In the case of a significant
inclination one would expect in the latter positions velocities closer
to the systemic velocity. The effect cannot expected to be easily seen in
the IC 5249 measurements, since the data are noisy and we can only take
positions about half a beam away from the disk. However, we see
absolutely no such effect.

\section{Observations of the stellar kinematics}

A spectrum was obtained with the Double Beam Spectropgraph at the
2.3 meter telescope of the Mount Stromlo and Siding Spring Observatories, 
located at Siding Spring. In total 8 separate exposures of 2000 seconds
each were obtained on April 15 and 16 and May 14 and 16, 1999. The spectrograph
slit was oriented along the major axis of IC 5249. The blue arm 
was set to register the wavelength range 4785 to 5752\AA, which would
include the Mgb-triplet and Fe-lines such as at 5270\AA, 
and the red arm the range 6005 to 6958\AA, which 
would include H$\alpha $ and the [NII] and [SII] lines. After logarithmic
binning of the data the pixels were 31.61 km/s in the blue arm and 25.28 km/s
in the red arm (original pixels were about 0.5\AA). As template star for 
the stellar velocity measurements HD132667 was used. 
The reduction took place along the lines of the crosscorrelation
analysis method of Statler (1995). Tests with broadening of the spectrum
of HD132667 and adding noise according to a signal-to-noise ratio of
15 showed that on our data the method worked reliably down to a velocity
dispersion of 20 km/s.

\begin{figure}
{\resizebox{\hsize}{!}{\includegraphics{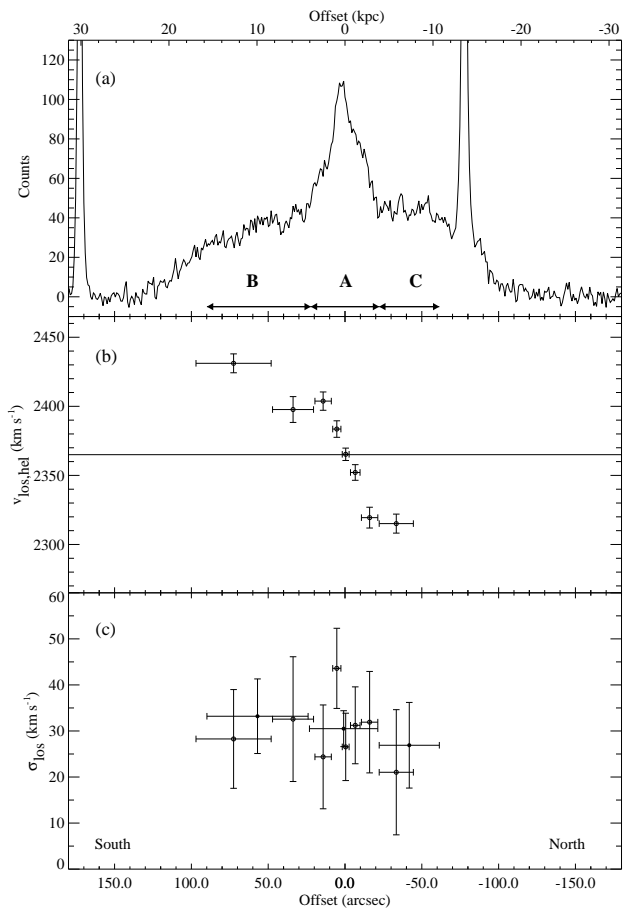}}}
\caption{Measurements of the stellar kinematics in IC 5249 from a long
slit spectrum along the major axis. The top panel shows the number of
counts along the slit (south is on the left side); 
the horizontal arrows show the ranges along the 
slit that are referred to in the text as (from left to right) B, A and
C. Note that two stars are on the slit of which the one on the right
can be identified in Fig.~\ref{picture}. 
The middle panel shows the observed stellar
radial velocity. The horizontal bars are  the ranges along the slit
over which data have been added to decrease the signal to noise. The
lower panel shows the velocity dispersions. Before adding data over
ranges along the slit these were shifted according to the run of radial
velocity in the middle panel.
}
\label{stellarvel}
\end{figure}

The basic results are in Fig.~\ref{stellarvel}. The data have been binned
along the slit before crosscorrelation until a signal-to-noise ratio of
15 was reached. The central bins were re-analysed after allowing for the
gradient in velocity (0.8 velocity pixels per spatial pixel) in order
to make the velocity dispersion free of radial velocity gradients accross
the final bin. Fig.~\ref{stellarvel} shows clear effects of rotation with
a flattening of the apparent rotation at about 70 km/s at 40 - 50\arcsec\  
(8 kpc). The center from which the offset has been measured is the brightest
spot along the disk (the possible bulge). The 
observed velocity dispersions are
consistent with a value of $\sim $30 km/s at all radii covered.

In order to improve the signal-to-noise further the data were binned in only
three ranges of distance from the presumable center. Over the region A,
roughly the central spike or bulge, from offsets -23\arcsec\ to +23\arcsec\ 
($\pm $4 kpc) the velocity dispersion (after allowing for the change in
radial velocity along this interval) was $30 \pm 4$ km/s (formal error).
For region B (to the south between -23\arcsec\ and -90\arcsec\ or 4 to
16 kpc) the velocity dispersion was $33 \pm 8$ km/s and on the corresponding
northern side (region C between +23\arcsec\ and +63\arcsec\ or 4 to 11 kpc)
the velocity dispersion came out as $26 \pm 9$ km/s. Taking the regions B and
C together (and shifting to the same radial velocity before adding) gives
a velocity dispersion of $33 \pm 5$ km/s.

Before we can use these data we need to concern ourselves first with two
line-of-sight (l.o.s.) effects. The first is absorption along the l.o.s.,
which would prevent us from seeing the full range of possible radial
velocities, which we will need to assume for our correction of the observed
velocity dispersions for l.o.s. integration. This is the reason why we
took the H$\alpha $ spectrum on the red arm. This spectrum showed H$\alpha $, 
[NII] and [SII] all along the disk with a few bright spots. In particular
the central area (around the possible bulge) is relatively bright in emission
lines.

\begin{figure}
{\resizebox{\hsize}{!}{\includegraphics{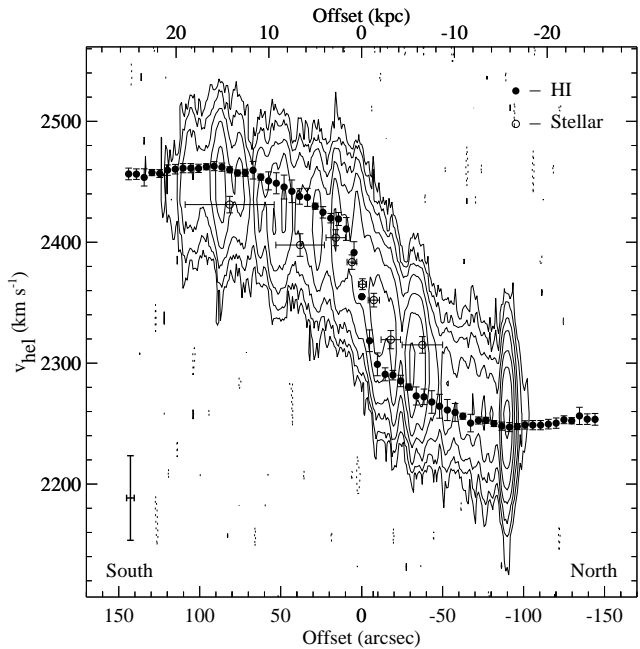}}}
\caption{The H$\alpha $-line (position versus velocity) along the major axis
of IC 5249 (contours). Superimposed on the H$\alpha $ data is 
the HI rotation curve of Fig.~\ref{rotcurve} and the stellar radial
velocities from the middle panel of Fig.~\ref{stellarvel}.
}
\label{Halpha}
\end{figure}

Fig.~\ref{Halpha} shows the H$\alpha $ line along the slit. Superimposed 
on this is the HI rotation curve from Fig.~\ref{rotcurve}. It is obvious that
we see the full amplitude of the HI rotation curve in the H$\alpha $ line
as well. This means that we can at these wavelengths (H$\alpha $
as well as the MgH triplet -- our 
stellar spectra are not too far apart in wavelength) we are looking at least
to the point where the full amplitude of the rotation comes along the line
of sight, or in other words {\it at least} (but presumably further 
than) half way through the galaxy. We also note that from the absence of
a measurable color change with $z$ in his photometry, Byun (1998) concludes
that there is little absorption in the plane of IC 5249.
In particular, Byun noted the absense of a dust
lane in photographs or photometry (a dip in vertical profiles). Wainscoat
(1983) also concluded from his near-infrared observations and the agreement 
with optical photometry that the effects of dust absorption in the optical
are small.

The H$\alpha $ line in Fig~\ref{Halpha} has an observed dispersion 
ranging from about 30 km s$^{-1}$ in the center to about 15 km s$^{-1}$
at the edges. This is similar to the HI data and is therefore
probably dominated by line of sight effects. It is likely that the
H$\alpha $ and HI velocity dispersions are similar, but the resolution
of the H$\alpha $ data is too coarse to do an analysis similar to that
on the HI data and determine the H$\alpha $ velocity dispersion reliably.

Having established that dust absorption is probably not a serious problem, 
we can consider the effects of the line of 
sight integration on the derived stellar radial velocities and dispersions.
This has been investigated for an average galaxy by Bottema et al. (1987)
and the effect was found to be small. The reason for that is that along
the path, the parts closest to the center contribute most
to the integral along the line of sight and the observed values are therefore
quite typical for what applies to these parts (both radial velocity and
velocity dispersion). However, in IC 5249 the situation might be different.
This is so, because the scalelength is very long and the truncation radius
only a few scalelengths, so that the contribution along the line of sight is 
decreasing only slowly while a smaller and smaller component of the rotation
comes into the line of sight.

The effect has been simulated for the following case: distance from
the center 8.5 kpc (50\arcsec), scalelength 11 kpc (the one used by Abe et 
al.), truncation radius 17 kpc, velocity dispersion constant along the line
of sight (this is likely in view of the large scalelength) and mean stellar
tangential velocity 100 km/s. Then it is found
that for the range 20 to 30 km/s for the velocity dispersion along the line
of sight the observed velocity dispersion is about 5 km/s lower than the
actual one at that radius from the center and the
mean tangential stellar velocity about 15 to 20 km/s lower. 
The first correction is comparable to the error, but the one in tangential 
velocity is substantial.

We therefore conclude  that the stellar rotation curve rises from the center
to a value of about 90-100 km/s at 50\arcsec, and that the (tangential)
velocity dispersion is within the uncertainties everywhere 25-30 km/s.

\section{The parameters of IC 5249}

\subsection{Is the central light concentration a small bulge?}

We first need to discuss whether or not the central part of the image of
IC 5249 constitutes a small bulge or a structure in the disk. Probably the most
important argument against it being a bulge, is that the $z$-profile
of the galaxy at this position is indistinguishable from those
at other radii and well-described by an exponential (except at small $z$).
Exponential profiles for bulges along their minor axes are not
uncommon (Andredakis \&\ Sanders, 1994), but it is not clear what a bulge
means in practice when its vertical distribution is the same as that of
the rest of the disk. A bulge is indicated when the observed central surface
brightness is well above that of the extrapolated exponential disk. In IC 
5249 this argument cannot be used, since the disk extends over only a few
scalelengths so that both fits with or without a bulge in this sense
are possible. The stellar velocity dispersion measured at the central
position is no different from what we find elsewhere in the disk as would
be expected if the central bright area is part of the disk.

At the central position there is an increase in the brightness of the
emission lines, but that could also be the result of the longer line
of sight through the galaxy. We conclude that the central bright concentration
is certainly not a bulge in the classical sense of a Population II
component and most likely
represents structure in the disk. One possibility would be
a bar, but we have no way at this stage to address that possibility with
more confidence.

\subsection{The scaleheight, scalelength and truncation radius of the stellar
disk}

For our analysis we first need to decide on the best values for the scale
parameters of the disk and its radius.

Byuan (1998) measured the scaleheight $h_{\rm z}$ in three color bands 
and found from his fits (after correcting to the distance used in 
this paper) $0.70 \pm 0.06$ kpc in $B$, $0.75 \pm 0.07$ kpc in 
$R$ and $0.71 \pm 0.08$ kpc in $I$. The Abe et al. (1999) photometry was
performed in the $R_{C}$- and $I_{C}$-bands. They finally arrive at a value
of $0.60 \pm 0.04$ kpc. Both sets of data show no evidence whatsoever for
a change in the scaleheight with distance from the center (although Byun
does have some evidence for a flaring just before the edge). For the analysis
below we will use a value of $h_{\rm z} = 0.65 \pm 0.05$ kpc.

The scalelength $h_{\rm R}$ is more problematic, since the outcome depends
on whether the central bright concentration is ignored (if it is a bulge)
or not (if it is part of the structure in the disk). The first possibility
is taken by Abe et al. and by Byun and they arrive respectively at values for
$h_{\rm R}$ of $11 \pm 2$ kpc and 13 kpc. If we would include the light
near the center a more appropriate value would be $h_{\rm R} = 7 \pm 1$ kpc 
(as adopted in van der Kruit, 1999). We will choose here for the latter
value on the basis of our discussion above.

The possibility should be considered that IC 5249 is not perfectly
edge-on and that this is the cause of the relatively large stellar
scaleheight and thickness of the HI layer. An inclination of
5$^{\circ }$ or so would project the stellar disk scalelength of 7 kpc onto an
apparent scaleheight as observed on the minor axis. 
The problem with that possibility is
that the radial HI distribution of Fig.~\ref{radial HI} would then result in an
apparent HI thickness (Fig.~\ref{thickness}) 
that is larger at the center than near
the edges, while the opposite is observed. Also for the optical disk the
apparent scaleheight should then decrease from the center outward and this is
also not observed. If the vertical light distribution is dominated by
inclination effects, that is if  the inclination would be of order 5$^{\circ }$,
then all contours (even the brightests) would have axis ratios 1:10. 
But the bright contours have an axis ratio of 1:20 or so. 
At smaller inclinations the observed stellar disk
scaleheight is not affected by the projected scalelength (see also van der
Kruit \&\ Searle, 1981a, their Fig. 3a).

There is little question that the observed radial surface brightness
profile shows a clear and sharp truncation in all measurements. The northern
side is more pronounced, but there is no doubt that the truncation
is symmetric in location. The radius is $R_{\rm max} = 17 \pm 1$ kpc.

We can compare these results to other edge-on galaxies, 
for example the sample
of de Grijs (1998), recently re-analysed by Kregel et al. (2001). First we
note that a scaleheight $h_{\rm z}$ of 0.65 kpc is somewhat, 
but not excessively 
high for a galaxy with a rotation velocity of about 100 km/s. However, the
radial scalelength $h_{\rm R}$ 
of 7 kpc is about twice what is found for such systems
in the de Grijs sample (and 11 or 13 kpc would be even more extraordinary).
The value for $h_{\rm R}/h_{\rm z}$ of about 11 does occur in other systems,
but those all have higher rotation velocities. For a $V_{\rm rot}$ of about 
100 km/s the usual value is about 6 or 7, although a few systems have
ratios closer to 10. The conclusion is that in terms of flattening 
$h_{\rm R}/h_{\rm z}$, IC 5249 is somewhat unusual in being flatter than 
galaxies with the same rotation velocities.

The truncation radius $R_{\rm max}$ gives $R_{\rm max}/h_{\rm R} = 2.4$. 
This is very low compared to other edge-on galaxies. The mean values found
by Schwarzkopf \&\ Dettmar (2000) and Kregel et al. (2001) are of the
order of $3.6 \pm 0.6$ (for a review see van der Kruit, 2001). Comparison
of the structural parameters in samples of edge-on galaxies (Kregel et
al., 2001) show that for its rotation velocity IC 5249 has reasonably 
normal $h_{\rm z}$ and $R_{\rm max}$, but an anomalously large $h_{\rm R}$. 

\section{Dynamics of IC 5249}

Our interest in this section will be to calculate various dynamical
parameters from the observed stellar
kinematical data (tangential velocity and velocity dispersion) and the HI
data (HI rotation curve, the HI surface density and the HI layer
thickness). The galactocentric distance where we have most
complete information is at about 40\arcsec\ (7 kpc, or about 1
radial scalelength). The other region, where
we have information is around 100\arcsec\ (17 kpc, near the optical
truncation). These regions coincide with radial peaks in the HI distribution,
ensuring that we have the most accurate information on the HI rotation
curve and layer thickness.

First we look at $R = 7$ kpc. The HI rotation curve is still rising 
there. We take $V_{\rm rot} = 90 \pm 5$ km/s and $dV_{\rm rot}/dR = 3 \pm 1$
km/s kpc$^{-1}$. Then the local Oort constants $A$ and $B$ are
\begin{equation}
A \equiv {1 \over 2} \left( {V_{\rm rot} \over R} - {{dV_{\rm rot}} \over dR}
\right) = 4.9 \pm 1.7 \ {\rm km\ s}^{-1}{\rm kpc}^{-1},
\end{equation}
\begin{equation}
B \equiv -{1 \over 2} \left( {V_{\rm rot} \over R} + {{dV_{\rm rot}} \over dR}
\right) = -7.9 \pm 2.7 \ {\rm km\ s}^{-1}{\rm kpc}^{-1},
\end{equation}
and the epicyclic frequency $\kappa $ is 
\begin{equation}
\kappa \equiv 2 \sqrt{B (B - A)} = 20 \pm 6 \ {\rm km\ s}^{-1}{\rm kpc}^{-1}.
\end{equation}
The epicyclic period is then $(3 \pm 1) \times 10^{8}$ years compared
to a rotation period of about $5 \times 10^{8}$ years, or more if there
is a significant asymmetric drift.

The ratio of the tangential versus radial velocity dispersion becomes
\begin{equation}
{\sigma _{\theta } \over \sigma _{\rm R}} = \sqrt{{- B} \over {A - B}} = 
0.79 \pm 0.07.
\end{equation}
Our observed velocity dispersion of $25 - 30$ km/s is mostly in the tangential
direction, so we estimate $\sigma _{\rm R} = 35 \pm 5$ km/s.

Now, let us first check that this is consistent with the asymmetric drift
equation for an exponential density distribution
\begin{eqnarray}
V_{\rm rot}^{2} - V_{\rm tan}^{2} &\approx &2 V_{\rm rot} (V_{\rm rot} -
V_{\rm tan}) \\
&= &\sigma _{\rm R}^{2} \left\{ {R \over h_{\rm R}} -
R {\partial \over {\partial R}} {\rm ln}\  \sigma _{\rm R}^{2} - 
\left[ 1 - { B \over {B - A}} \right] \right\}, \nonumber
\end{eqnarray}
where $V_{\rm tan}$ is the tangential velocity of the stars. We don't know
the second term in the righthand part, but it is often thought that it is
of order $-R/h_{\rm R}$ (van der Kruit \&\ Freeman, 1986).
With our values this equation gives an asymmetric drift 
$V_{\rm rot} - V_{\rm tan} = 10
\pm 3$ km/s, which is consistent with our measurements.

The values that we find for the thickness of the HI layer (the 
Gaussian dispersion) and the stellar exponential scaleheight are 
comparable. At face value we would have to conclude that the gas-layer is 
thicker than the stellar disk; 
this is difficult to fit into any scenario that we can imagine. 
We will first assume that the vertical distribution of the gas
and the stars is the same. Then both gas and stars will have the same (vertical)
velocity dispersion, so that very little dynamical evolution
can have occured in the stellar disk at least {\it in the $z$-direction}.

A general case for the $z$-distribution of matter that is useful for
our treatment is (van der Kruit, 1988)
\begin{equation}
\rho (z) = \rho _{\circ} {\rm sech}^{2/n} \left( {{n z} \over
{2 h_{\rm z}}} \right).
\end{equation} 
De Grijs et al. (1997) have found from fits to actual (edge-on) galaxies
that
\begin{equation}
{2 \over n} = 0.54 \pm 0.20.
\end{equation}
Now if we allow for the range of density laws from $2/n = 1$ (a 
sech$(z/h_{\rm z})$ dependence) 
to the exponential case $2/n = 2/\infty $ (an exp$(-z/h_{\rm z})$ 
dependence) and integrate
the vertical velocity dispersion (weighed by the density) we obtain
(van der Kruit, 1988, eq. (23) - (25))
\begin{equation}
\sigma ^{2}_{\rm z} = (5.0 \pm 0.3) G \Sigma (R) h_{\rm z},
\end{equation} 
where $\Sigma (R)$ is the disk surface density at radius $R$ and $G$ the
gravitational constant. If the stars and the gas
have the same vertical distribution, they should also have the same
velocity dispersion. So we take $\sigma _{\rm z} = 7 - 8$ km/s,
which then gives with $h_{\rm z} = 0.65 \pm 0.05$ kpc
\begin{equation}
\Sigma (7\ {\rm kpc}) = 4.0 \pm 0.7 \ {\rm M}_{\odot}{\rm pc}^{-2}.
\end{equation}
From Fig.~\ref{radial HI}, we see that at the relevant radius the HI surface
density is of order 3.2 M$_{\odot }$ pc$^{-2}$ ($= 4.0 \times 10^{20}$ HI
atoms cm$^{-2}$). However, adding another third
to allow for helium and keeping in mind that in the observations ATCA
missed probably a not insignificant amount of the 21-cm line flux, we see
that this mass surface density estimate is already accounted for by the
gas alone. 

Abe et al. fit the disk with a radial scalelength of 11 kpc and for that fit
quote a central, face-on surface brightness for IC 5249 of 24.3 mag
arcsec$^{-2}$ in the $R_{\rm C}$-band. This corresponds to 4.3 L$_{\odot}$
pc$^{-2}$. In our preferred fit with $h_{\rm R}$ = 7 kpc, the central value
has become 7.6 L$_{\odot }$ pc$^{-2}$. In both cases at $R = 7$ kpc the surface
brightness is about 2.5 L$_{\odot }$ pc$^{-2}$. For any reasonable
mass-to-light ratio we expect a few M$_{\odot }$ pc$^{-2}$ in stars.

So we see that we do not get a consistent result if indeed the HI disk
is resolved by the ATCA observations. 
However, if the apparent thickness of the HI layer is affected by the
inclination, the HI layer must be thinner than the stellar disk and the
value derived for the surface density becomes a lower limit. This seems
to make more sense.

At this point it is useful to set an upper limit on the disk mass surface
density from the rotation curve. We use the simple case of
an self-gravitating exponential disk, for which the rotation curve reaches
a maximum at about 2.2$h_{\rm R}$ of
\begin{equation}
V^{\rm max}_{\rm rot} = 0.88 \sqrt{\pi G \Sigma (0) h_{\rm R}},
\end{equation}
(Freeman, 1970)
and require that this is at most equal to the observed maximum rotation of
$105 \pm 5$ km/s. Then the upper limit on the central surface density is
\begin{equation}
\Sigma (0) \le 150 \pm 20\  {\rm M}_{\odot}{\rm pc}^{-2}.
\end{equation}
At $R = 7$ kpc or one scalelength this gives an upper limit 
\begin{equation}
\Sigma(7\ {\rm kpc}) \le 55 \pm 7\  {\rm M}_{\odot}{\rm pc}^{-2}. 
\end{equation}
Further estimates of the surface densities can be made in two ways. First,
we can use the global stability criterion of Efstathiou et al. (1982)
\begin{equation}
Y \equiv V_{\rm max} \left( {h \over {G M_{\rm D}}} \right) ^{1/2} \sim 1.1
\end{equation}
to make another estimate of the disk mass. In this criterion an exponential
disk is assumed with mass $M_{\rm D}$, which is stabilised by a dark halo
($V_{\rm max}$ is the maximum velocity in the rotation curve). We then find
\begin{equation}
{\rm M}_{\rm D} = (1.5 \pm 0.2) \times 10^{10} \ {\rm M}_{\odot},
\end{equation}
and taking account of the truncation radius this translates into
\begin{equation}
\Sigma (0) = (70 \pm 9)\ {\rm M}_{\odot}{\rm pc}^{-2},
\end{equation}
\begin{equation}
\Sigma (7\ {\rm kpc}) = (26 \pm 3)\ {\rm M}_{\odot}{\rm pc}^{-2}.
\end{equation}
Secondly, we can use the local stability criterion 
$Q$\lower.5ex\hbox{$\; \buildrel > \over \sim \;$}1 of Toomre (1964) to estimate
the disk surface density by rewriting the definition of $Q$ as
\begin{equation}
\Sigma (R) = {{\sigma _{\rm R} \kappa } \over {3.36 G Q}},
\end{equation}
where we can assume that $Q \approx 2$ (Bottema, 1993, 1997; van der Kruit,
1999). Doing this with our values we get
\begin{equation}
\Sigma (7\ {\rm kpc}) = (25 \pm 8)\ {\rm M}_{\odot}{\rm pc}^{-2}.
\end{equation}
On the basis of the above we adopt at $R \sim 7$ kpc a
surface density of 25 M$_{\odot}$pc$^{-2}$ (which would then be
dominated by the stars). This would imply a vertical stellar 
velocity dispersion of
\begin{equation}
\sigma _{\rm z} = \sqrt{(5.0 \pm 0.3) G \Sigma (R) h_{\rm z}} = 19 \pm 4\ {\rm
km}\ {\rm s}^{-1}.
\end{equation}
We can then also use the relation between the FWHM of the HI layer (van
der Kruit, 1981) and the stellar disk parameters
\begin{equation}
({\rm FWHM})_{\rm HI} = (2.8 \pm 0.2) \sigma _{\rm HI} \sqrt{ h_{\rm z} \over
{2 \pi G \Sigma (R)}},
\end{equation}
where the constant again has been adapted to allow for a density distribution
between a sech$(z/h_{\rm z})$ and an exponential exp$(-z/h_{\rm z})$
dependence. Then
\begin{equation}
({\rm FWHM})_{\rm HI} = 0.60 \pm 0.17\ {\rm kpc}.
\end{equation}

We repeat this analysis for the radius $R$ = 17 kpc. There are two important
changes here. Following Byun (1998) we see that there is a possibility
that at this radius there is a sudden flaring of the stellar disk with
the scaleheight $h_{\rm z}$ increasing to about $1.0 \pm 0.1$ kpc. Secondly
the HI disk might here not be affected by inclination effects (for an
illustration of this see the simulations in van der Kruit, 1981, especially
his Fig. 6) and it is
a real possibility that the HI layer and the stellar disk do have the
same thickness. We now observe a essentially flat rotation curve and adopt
$V_{\rm rot} = 105 \pm 5$ km/s and $dV_{\rm rot}/dR = 0 \pm 1$ km/s 
kpc$^{-1}$. 

The dynamical properties then become at $R$ = 17 kpc:
\begin{equation}
A = - B = 3.1 \pm 0.6\ {\rm km\ s}^{-1}{\rm kpc}^{-1},
\end{equation}
\begin{equation}
\kappa = 3.5 \pm 0.7\ {\rm km\ s}^{-1}{\rm kpc}^{-1}.
\end{equation}
The period in the epicycle now is $(1.8 \pm 0.4) \times 10^{9}$ years,
compared to a rotation period of about $1 \times 10^{9}$ years.
\begin{equation}
{\sigma _{\theta } \over \sigma_{\rm R}} = 0.71 \pm 0.10.
\end{equation}
We cannot calculate here the asymmetric drift.
Assuming that the gas and the stars have the same vertical distribution
we find
\begin{equation}
\Sigma (17\ {\rm kpc}) = 2.3 \pm 0.5\ {\rm M}_{\odot }{\rm pc}^{-2}.
\end{equation}
The observed surface density of HI is about 4.8 M$_{\odot }$pc$^{-2}$,
again showing an inconsistency if the gas and the stars have a similar 
$z$-distribution.

For the maximum disk estimate we now get
\begin{equation}
\Sigma (17\ {\rm kpc}) \le 13 \pm 2\ {\rm M}_{\odot }{\rm pc}^{-2},
\end{equation}
and from the global stability parameter
\begin{equation}
\Sigma (17\ {\rm kpc}) = 6 \pm 1\ {\rm M}_{\odot }{\rm pc}^{-2}.
\end{equation}
We cannot use the local stability criterion, since we do not know the
stellar radial velocity dispersion. However, above we adopted a surface 
density of 25 M$_{\odot }$pc$^{-2}$ at $R$=7 kpc, and if our exponential
disk has a constant mass-to-light ratio this also implies a surface
density at $R$=17 kpc of about 6 M$_{\odot }$pc$^{-2}$.
Taking a surface density of 6 M$_{\odot }$pc$^{-2}$, we get
\begin{equation}
\sigma _{\rm R} = 25 \pm 5\ {\rm km\ s}^{-1},
\end{equation}
\begin{equation}
\sigma _{\rm z} = 11 \pm 2\ {\rm km\ s}^{-1},
\end{equation}
\begin{equation}
({\rm FWHM})_{\rm HI} = 1.5 \pm 0.5\ {\rm kpc}.
\end{equation}

So, we conclude from this section that the stellar radial velocity dispersion
at R = 7 kpc is about 35 km/s from actual measurements and we infer
that the vertical velocity dispersion is about 20 km/s. Near the disk edge
we infer a radial velocity dispersion of about 25 km/s and
a vertical one of about 10 km/s. The observed thickness of the HI layer
in the total HI map is a result of a small deviation of the plane of
IC 5249 away from perfectly edge-on. 

\section{Discussion}

From the photometric data on IC 5249 is it clear that its very thin appearance
on the sky is the result of a combination of a number of circumstances.
These are the fact that the galaxy has a low surface brightness disk (when
seen face-on) plus the rather long scalelength and the sharp truncation
at only a little more than 2 scalelengths. As a result of
this only the peak in the $z$-profiles reaches above sky brightness and
this peak varies then little with distance from the center before ending
abruptly.

Remarkable is that in spite of its special nature, IC 5249 falls close to
the relations that Bottema (1993) found between a fiducial velocity dispersion
(here the radial one at one scalelength, which we found to be about 35 km/s)
and the integrated luminosity of the disk and the maximum rotation velocity.
These Bottema relations are most easily understood, if the product
of central face-on surface brightness, mass-to-light ratio and Toomre $Q$ are
constant between galaxies (see e.g. van der Kruit \&\ de Grijs, 1999). 
The derivation of that result makes use of the
Tully-Fisher relation (it needs to have an exponent 4).

We already noted that in terms of flattening ($h_{\rm R}/h_{\rm z}$), IC 5249
is slightly unusual compared to the sample studied in Kregel et al. (2001), 
where at a rotation velocity of 100 km/s this ratio ranges between 7 and 10.
For IC 5249 it is just above 10 for our scalelength of 7 kpc, but would be
larger if the central bright region really is a bulge and the scalelength is
more like 11 or 13 kpc. Both the scalelength and in particular the scaleheight
are significantly larger than in classical edge-on galaxies as in van der
Kruit \&\ Searle (1981, 1982) or our Galaxy.

It is interesting to note that the stellar velocity dispersions at $R$=7 kpc
($\sigma _{\rm R} \sim $ 35 km/s, $\sigma _{\theta} \sim $ 30 km/s and 
$\sigma _{\rm z} \sim $ 20 km/s) are comparable to those in the solar 
neighborhood, as reported by Dehnen \&\ Binney (1998) for the redder stars
(and therefore the old disk population). This means that in IC 5249
with a smaller surface density as much dynamical evolution has occured
as in the Galaxy.
It seems that in IC 5249 the star formation has proceeded slower: as far
as we can tell a relatively large fraction of the disk mass is still in the 
form of gas (at least 3.2 M$_{\odot}$ pc$^{-2}$ in HI depending on where
the HI missed by the interferometer is, multiplied by 1.33 to allow for helium
and some more ionized and possibly molecular gas, compared to our adopted
surface density of 25 M$_{\odot}$ pc$^{-2}$). 

While at $R$=7 kpc the axis ratio of the velocity ellipsoid 
($\sigma _{\rm R}/\sigma _{\rm z}$)
is comparable to the solar neightborhood, it seems to be more extreme at the
outer boundaries of the disk. These are inferred values and should
be treated with caution.

\begin{table}
\caption[]{Comparison between IC 5249 and UGC 7321}
\label{comptable}
\[
\begin{tabular}{l l l}
\hline
\noalign{\smallskip}
 &  IC 5249 & UGC 7321 \\
\noalign{\smallskip}
\hline
\noalign{\smallskip}
$h_{\rm R}$ (kpc) & 7 & 2.1 \\
$h_{\rm z}$ (kpc) & 0.65 & 0.15 \\
$h_{\rm R}/h_{\rm z}$ & 11 & 14 \\
$R_{\rm max}$ (kpc) & 17 & 7.3 \\
$R_{\rm max}/h_{\rm R}$ & 2.4 & 3.5 \\
$\mu _{\circ }$ (B-mag arcsec$^{-2}$) & $\sim $ 24.5 & $\sim $ 23.4 \\
$L_{\rm tot,B}$ (L$_{\odot }$) & $\sim 6 \times 10^{9}$ & $\sim 9 \times 
10^{8}$ \\
$V_{\rm rot}$ (km/s) & 105 & 106 \\
$M_{\rm HI}$ (M$_{\odot}$) & $8.5 \times 10^{9}$ & $1.1 \times 10^{9}$ \\
$M_{\rm disk}$ (M$_{\odot }$) & $\sim 1.5 \times 10^{10}$ & $\sim 5 \times 
10^{9}$ \\
\noalign{\smallskip}
\hline
\end{tabular}
\]
\end{table}

In Table~\ref{comptable} we compare some properties of IC 5249 and UGC 7321,
where for the latter we took the analysis of Matthews et al. (1999) and 
Matthews (2000). The disk mass is the one estimated from the global
stability criterion, which was also done by Matthews. We notice that
these two ``superthin'' galaxies have --apart from this appearance 
and the fact that face-on they would be low surface brightness
galaxies-- very little
in common; in fact the only  remarkable similarity is the rotation
velocity. In fact UGC 7321 is indeed very thin is absolute measure, while
IC 5249 should in that respect be seen as relatively thick. However,
UGC 7321 is a small galaxy in the sense of the linear size, while IC 5249
is not. 
Matthews (2000), using quite similar arguments as here, estimates that at one
scalelength the vertical velocity dispersion of the stars is only of the order
of 12 km/s or so. This is larger than that of cold HI and relatively
little dynamical evolution must have occured.
This is not the case in IC 5249. It would be of much interest to measure
the radial stellar velocity dispersion in UGC 7321. 

In conclusion, we have shown that IC 5249 is a large linear size low surface
brightness galaxy in which the dynamical evolution of the velocity dispersions
of the stars --at one scalelength from the center--
has proceeded to similar values as in the solar neighborhood. Apparently,
the angular momentum distribution of the gas that formed the disk in
IC 5249 was such that, compared to the Galaxy, 
a much more extended distribution resulted in spite of
the lower overall rotation and mass. The low surface density that arose from 
that gave rise to a thicker HI layer in which star formation proceeded at
a much slower rate (a considerable fraction of the disk mass is probably
still in the form of gas). In spite of that, disk heating proceeded
at a similar pace. It is possible that this dynamical evolution resulted
from strong (in a gravitational sense) spiral structure and it would be
interesting to know what IC 5249 would look like face-on. We have not
been able to find the means or travel funds to perform that observation.

\begin{acknowledgements}
We thank the Australia Telescope Compact Array archive for providing us with
the HI observations of IC 5249. We are grateful to the director of
Mount Stromlo and Siding Spring Observatories for observing time on the 
2.3 meter telescope. PCK thanks the Space Telescope Science Institute for 
hospitality and support during a period when most of this paper was 
written. JJ-V would like to thank R. Bottema and I. Garcia-Ruiz for their
help with the reduction and analysis of the radio data.
The Area Board for Exact Sciences of the Netherlands
Organisation for Scientific Research (NWO) supported the travel of PCK and
(partly) MK. 
\end{acknowledgements}


\begin{thebibliography}{}

\bibitem[]{}
Abe, F., Bond, I.A., Carter, B.S., et al., 1999, AJ 118, 261

\bibitem[]{}
Andredakis, Y.C., Sanders, R.H., 1994, MNRAS 267, 283

\bibitem[]{}
Bottema, R., 1993, A\&A 275, 16

\bibitem[]{}
Bottema, R., 1997, A\&A 328, 517

\bibitem[]{}
Bottema, R., van der Kruit, P.C., Freeman, K.C., 1987, A\&A 178, 77

\bibitem[]{}
Byun, Y.-I., 1998, Chin. J. Phys. 36, 677

\bibitem[]{}
Carignan, C., 1983, Ph.D. Thesis, Australian National University

\bibitem[]{}
de Grijs, R., 1998, MNRAS 299, 595

\bibitem[]{}
de Grijs, R., Peletier, R.F., van der Kruit, P.C., 1997, A\&A 327,
966

\bibitem[]{}
Dehnen, W., Binney, J.J., 1998, MNRAS 298, 387

\bibitem[]{}
Efstathiou, G., Lake, G., Negroponte, J., 1982, MNRAS 199, 1069

\bibitem[]{}
Freeman, K.C., 1970, ApJ 160, 811

\bibitem[]{}
Gilmore, G., Reid, N., 1983, MNRAS 202, 1022

\bibitem[]{}
Jenkins, A., 1992, MNRAS 257, 620

\bibitem[]{}
Jenkins, A., Binney, J.J., 1990, MNRAS 245, 305

\bibitem[]{}
Kregel, M., van der Kruit, P.C., de Grijs, R. 2001, MNRAS (submitted)

\bibitem[]{}
Mathewson. D.S., Ford, V.L., Buchhorn, M., 1992, ApJS 81, 413

\bibitem[]{}
Matthews, L.D., 2000, AJ 120, 1764

\bibitem[]{}
Matthews, L.D., Gallagher III, J.S., van Driel, W., 1999, AJ 118, 2751

\bibitem[]{}
Mendelowitz, C., 2000, NROA Student Project, \\
http://www.cv.nrao.edu/$\sim $jhibbart/students/CMendelo\-witz/caylin.html

\bibitem[]{}
Schwarzkopf, U., Dettmar, R.-J., 2000, A\&A 361, 451

\bibitem[]{}
Statler, T.S., 1995, AJ 109, 1371

\bibitem[]{}
Toomre, A., 1964, ApJ 139, 1217

\bibitem[]{}
van der Kruit, P.C., 1981, A\&A 99, 298

\bibitem[]{}
van der Kruit, P.C., 1988, A\&A 192, 117

\bibitem[]{}
van der Kruit, P.C., 1999, Ap\&SS 269/270, 139

\bibitem[]{}
van der Kruit, P.C., 2000, in: The Legacy of J.C. Kapteyn, eds. P.C. van
der Kruit and K. van Berkel (Dordrecht: Kluwer), 299

\bibitem[]{}
van der Kruit, P.C., 2001, in: Galaxy Disks and Disk Galaxies, eds. J.G.
Funes S.J. and E.M. Corsini, ASP Conf. Series (in press).

\bibitem[]{}
van der Kruit, P.C., de Grijs, R., 1999, A\&A 352, 129

\bibitem[]{}
van der Kruit, P.C., Freeman, K.C., 1986, ApJ 303, 556

\bibitem[]{}
van der Kruit, P.C., Searle, L., 1981, A\&A 95, 105

\bibitem[]{}
van der Kruit, P.C., Searle, L., 1982, A\&A 110, 61

\bibitem[]{}
Wainscoat, R.J., 1986, Ph.D. Thesis, Australian National University



\end{thebibliography}
\end{document}